\documentclass[journal=jacsat,manuscript=article,sorting=none]{achemso}
\usepackage[version=3]{mhchem} 

\author{Carlos C. Soares}
\affiliation[FisMat]{Física de Materiais, Escola Politécnica de Pernambuco, UPE, Recife, Pernambuco 50720-001, Brazil} 

\email{carlos.soares@lnls.br}
\phone{+55 (21)99525-2535}

\alsoaffiliation[LNLS]{Laboratório Nacional de Luz Síncrotron, Centro Nacional de Pesquisa em Energia e Materiais, 13083-970, Campinas, SP, Brazil}
\alsoaffiliation[IFGW]{Instituto de Física Gleb Wataghin, Universidade Estadual de Campinas, Campinas, SP, 13083-859, Brazil}

\author{Thiago J. A. Mori}
\affiliation[LNLS]{Laboratório Nacional de Luz Síncrotron, Centro Nacional de Pesquisa em Energia e Materiais, 13083-970, Campinas, SP, Brazil}

\author{Fanny Béron}
\affiliation[IFGW]{Instituto de Física Gleb Wataghin, Universidade Estadual de Campinas, Campinas SP, 13083-859, Brazil}

\author{Jagadeesh S. Moodera}
\affiliation[MIT]{Plasma Science and Fusion Center, and Francis Bitter Magnet Laboratory, MIT, Cambridge, MA 02139, USA}
\alsoaffiliation[MIT]{Department of Physics, Massachusetts Institute of Technology, Cambridge, Massachusetts 02139, USA}

\author{Júlio C. Cezar}
\affiliation[LNLS]{Laboratório Nacional de Luz Síncrotron, Centro Nacional de Pesquisa em Energia e Materiais, 13083-970, Campinas, SP, Brazil}

\author{Jeovani Brandão}
\affiliation[LNLS]{Laboratório Nacional de Luz Síncrotron, Centro Nacional de Pesquisa em Energia e Materiais, 13083-970, Campinas, SP, Brazil}
\email{jeovani.brandao@lnls.br}

\author{Gilvânia Vilela}
\affiliation[FisMat]{Física de Materiais, Escola Politécnica de Pernambuco, UPE, Recife, Pernambuco 50720-001, Brazil}
\alsoaffiliation[MIT]{Plasma Science and Fusion Center, and Francis Bitter Magnet Laboratory, MIT, Cambridge, MA 02139, USA}
\email{gilvania.vilela@upe.br}

\title[An \textsf{achemso} demo]
  {Compensation-Like Temperature and Spin-Flip Switch in Strained Thulium Iron Garnet Thin Films: Tuning Sublattice Interactions for Ferrimagnetic Spintronics\footnote{Compensation-like temperature and spin-flip switch in strained Thulium Iron Garnet thin films: Tuning Sublattice Interactions for Ferrimagnetic Spintronics}
  }
   
\abbreviations{IR, NMR, UV}

\begin{document}


\begin{abstract}

Certain rare-earth iron garnet (RIG) thin films combine desirable properties such as low magnetic damping, high magnetostriction, and, in some cases, perpendicular magnetic anisotropy (PMA), making them attractive for spintronics applications. However, the interplay between their magnetic sublattices in confined films remains poorly explored, particularly the coupling between 3d and 4f electrons. Here, we investigate the magnetic properties of a 30 nm-thick thulium iron garnet (TmIG) thin film, where tensile strain promotes PMA. SQUID magnetometry and X-ray Magnetic Circular Dichroism measurements reveal a magnetization minimum near 50 K under moderate magnetic fields, leading to a compensation-like temperature (\(T_{\text{comp-like}}\)), a feature absent in bulk TmIG. The presence of \(T_{\text{comp-like}}\) is particularly relevant for controlling magnetization dynamics through compensation phenomena. Additionally, we observe a field-induced spin-flip transition in the Tm sublattice, where Tm moments reorient and align ferromagnetically concerning the Fe sublattices. This mechanism can be exploited for energy-efficient magnetization reversal. These findings provide new insights into strain-driven magnetic phenomena in rare-earth iron garnet thin films, highlighting the interplay between exchange interactions and anisotropy in confined geometries, which is crucial for the development of spintronic and magnonic devices.

\end{abstract}

\maketitle

\section{Introduction}

Rare-earth iron garnets (RIGs), with the general formula \(\mathrm{R}_3\mathrm{Fe}_5\mathrm{O}_{12}\) (R = Gd, Tb, Dy, Ho, Er, Tm, Yb, or Lu), are insulating ferrimagnets that have attracted considerable attention due to their rich sublattice structure and a combination of properties that can be tailored for spintronic and magnonic applications\cite{Duong22-Interfacial, Vilela20-Magnon, Strohm12-Element, Cornelissen15-Long}. Among these, specific RIG compositions exhibit low magnetic damping, elevated Curie temperatures ($T> 500$K), and the ability to support perpendicular magnetic anisotropy (PMA) under strain \cite{Zanjani19-Thin, Damerio23-Sputtered}. In particular, thulium iron garnet (TmIG, \(\mathrm{Tm}_3\mathrm{Fe}_5\mathrm{O}_{12}\)) stands out for its distinct magnetic properties, including a significantly higher magnetostriction constant (\(\lambda_{111} \approx -5.2 \times 10^{-6}\)), nearly twice that of yttrium iron garnet (YIG, \(\mathrm{Y}_3\mathrm{Fe}_5\mathrm{O}_{12}\)) \cite{Haubenreisser-Physics}. This enhanced magnetostriction facilitates the development of PMA under strain in TmIG thin films grown on Gadolinium Gallium Garnet (GGG) substrates, making it particularly promising for applications in magnetic sensors and memory devices, where PMA enhances storage density, thermal stability, and observability of spin-related phenomena\cite{Vilela20-Magnon, Zhang22-Strong}.

The magnetic structure of TmIG consists of three sublattices: tetrahedral (\(\mathrm{Fe}_{d}\)) and octahedral (\(\mathrm{Fe}_{a}\)) sites both occupied by \(\mathrm{Fe}^{3+}\) ions, completed by dodecahedral (\(\mathrm{Tm}_{c}\)) sites occupied by \(\mathrm{Tm}^{3+}\) ions. The Fe sublattices couple antiferromagnetically, resulting in a net magnetization dictated by the imbalance between their moments (Figure \ref{Fig1} (a)). At high temperatures ($T>200$ K), the iron sublattices dominate due to strong superexchange interactions \cite{Geller65-Magnetic, Bayarra-Tuning}. However, at low temperatures, the Tm sublattice becomes increasingly relevant, with its magnetic moment growing due to strong spin-orbit coupling and local anisotropy effects. In other rare-earth iron garnets, such as TbIG and ErIG, these interactions give rise to a compensation temperature (\(T_{\text{comp}}\)), a well-established feature in RIGs resulting from the competition between rare-earth and Fe sublattices. In some cases, particularly under strain or at low temperatures, this regime has also been associated with non-collinear spin configurations, such as Canted or umbrella-like structures \cite{Pickart70-RareEarth, Lahoubi14-Anomalous, Strohm12-Element}. In contrast, bulk TmIG does not exhibit compensation temperature, likely due to weaker indirect exchange coupling between the Tm and Fe sublattices \cite{Geller65-Magnetic}.

The presence of \(T_{\text{comp}}\) in ferrimagnetic systems is of high technological interest, as it allows for energy-efficient spin manipulation and ultrafast magnetization reversal, key properties for high-speed magnetic memory and spintronic devices \cite{Deb18-Controlling, Zhao16-Origin}. In rare-earth-based magnetic oxides, compensation phenomena are considered to enhance the efficiency of the magnetization switching, making materials with tunable \(T_{\text{comp}}\) promising candidates for the next generation of magnetic storage \cite{Cao14-Magnetization}. Furthermore, field-induced spin flips, commonly observed in rare-earth orthochromites, have been explored for controlled magnetization reversal under external excitations, such as electric and optical stimuli \cite{Cao14-Magnetization}. These effects demonstrate that rare-earth-based materials with compensation-like behavior and spin reorientation transitions provide a versatile platform for tailoring magnetic properties through external control.

Strain engineering in epitaxial thin films introduces additional complexity to the 3 {\it d} and 4{\it f} sublattices interactions. In rare-earth iron garnets such as TbIG, the interplay between local anisotropy effects, spin-orbit coupling, and exchange interactions led to complex spin configurations, including canted structures such as umbrella-like arrangements \cite{Lahoubi12-Symmetry, Lahoubi14-Anomalous, Tomasello22-Origin}. These non-collinear magnetic structures arise from the competition between crystal field effects and exchange interactions, particularly at low temperatures ($T<100$K) \cite{Lahoubi14-Anomalous, Tomasello22-Origin}. While such effects were observed in TbIG thin films an GGG substrate, their presence in TmIG thin films remains an open question. Although previous studies did not explicitly address strain, similar modifications in magnetic anisotropy may occur in TmIG thin films due to strain-induced effects. Previous research showed that strain plays a role in establishing a perpendicular magnetic anisotropy (PMA) in thin films of TmIG and Bismuth-doped TmIG \cite{Zhang22-Strong, Vilela20-Strain}. For example, a 30-nm-thick TmIG film over a GGG substrate annealed at 600°C exhibited compressive strain, favoring an in-plane anisotropy, whereas annealing above 800°C increased the out-of-plane lattice parameter, promoting a strong PMA \cite{Vilela20-Strain}. 

In this study, we investigated the magnetic properties of a 30-nm-thick TmIG thin film over a (111) oriented GGG substrate annealed at 900°C, a condition that induces tensile strain, creating a stable PMA \cite{Vilela20-Strain}. The SQUID magnetometry and element-specific X-ray Magnetic Circular Dichroism (XMCD) measurements revealed a distinct compensation-like temperature (\(T_{\text{comp-like}}\)) near 50 K under moderate magnetic fields, absent in bulk TmIG, which we attribute to modifications in exchange interactions and sublattice anisotropies. Additionally, we observed a field-induced spin-flip of the Tm sublattice, a phenomenon not observed for the Fe sublattices, highlighting the distinct response of the rare-earth and transition-metal sublattices to external fields. This study advances both the fundamental understanding of TmIG thin films and their technological potential for future spintronic and magnonic applications.

\section{Experiments}

\begin{figure*}[ht]
    \centering
    \includegraphics[width=\textwidth]{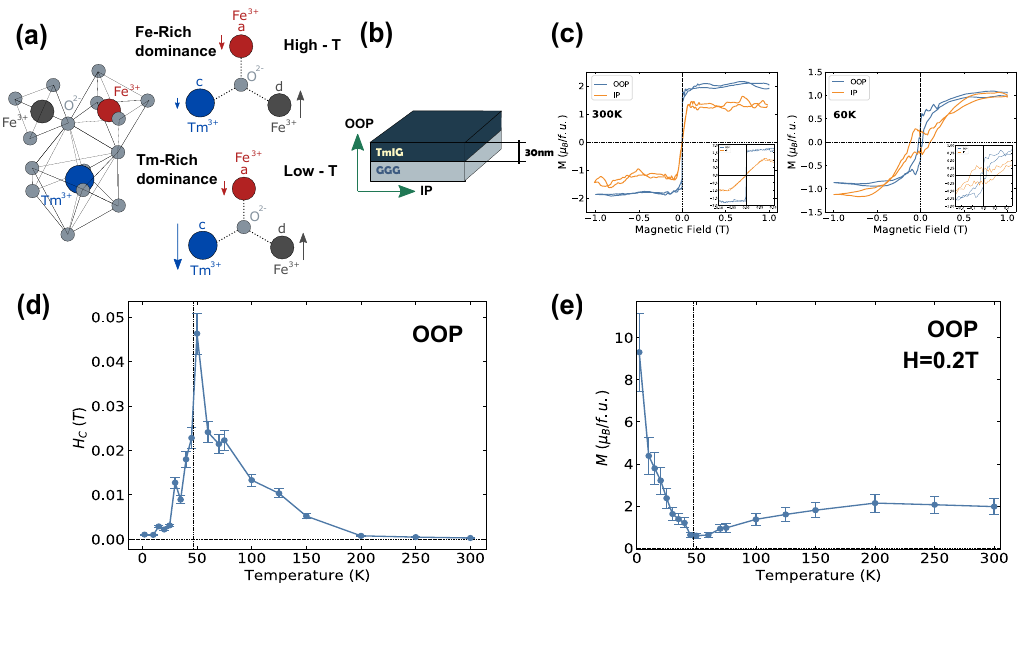}
    \caption{
    (a) Formula unit diagram of TmIG showing the sublattice structure along with respective magnetic interactions at high and low temperatures. The labels indicate the crystallographic symmetries of the magnetic ions: \(a\) for tetrahedral (Fe), \(d\) for octahedral (Fe), and \(c\) for dodecahedral (Tm). 
    (b) Schematic of the 30 nm-thick TmIG film grown on a GGG substrate, including the sample layout and magnetic measurement configurations for out-of-plane (OOP) and in-plane (IP) fields. 
    (c) Magnetic hysteresis loops measured at 300 K (left) and 60 K (right) for both out-of-plane (OOP) and in-plane (IP) magnetic field orientations. Insets provide a detailed view of the low-field regime at each temperature to emphasize subtle features in the magnetization behavior.
    (d) Coercive field (\(H_c\)) as a function of temperature, showing a divergence near 50 K at 1 T.
    (e) Magnetization (\(M\)) as a function of temperature at an applied magnetic field of 0.2 T.
    }
    \label{Fig1}
\end{figure*}

TmIG thin films were deposited on (111)-oriented GGG substrates. Before deposition, the substrates were annealed in a quartz tube furnace at 1000 °C for 6 h under an oxygen-enriched atmosphere of 1 atm pressure to enhance surface crystallinity. The deposition was performed in a UHV sputtering chamber with a base pressure below 5 × $10^{-8}$ Torr, using a 99.9\% pure TmIG target, an RF power of 50 W, and an argon working pressure of 2.8 mTorr. The films were grown at room temperature, followed by post-growth annealing at 900°C for a duration of 8 h in flowing oxygen. For more detailed information on the fabrication and characterization of TmIG thin films, refer to Ref.\cite{Vilela20-Strain}. A schematic of the sample is presented in Figure \ref{Fig1}(b).

Magnetic characterization was conducted in a temperature range from 2 K to 300 K using a Quantum Design MPMS3 superconducting quantum interference device (SQUID) magnetometer. The magnetic field was applied both perpendicular and parallel to the surface of the film, and hysteresis loops were measured at several temperatures. The paramagnetic contribution from the GGG substrate was subtracted to determine the TmIG magnetic moment at each temperature accurately.

To determine the site-specific magnetic contributions of each Fe and Tm sublattices, we both performed X-ray absorption spectroscopy (XAS) and X-ray magnetic circular dichroism measurements (XMCD) under various magnetic fields and temperatures. These experiments were performed at the Soft X-ray Absorption and Imaging (SABIÁ) beamline of the Brazilian Synchrotron Light Laboratory (LNLS). For Tm, the spectra were collected at the $M_{4,5}$ absorption edges, while for Fe, the measurements focused on the $L_{2,3}$ absorption edges. XMCD spectra were obtained by calculating the difference between XAS spectra acquired with right ($\mu_+$) and left ($\mu_-$) handed circularly polarized light. During the XMCD measurements, the samples were magnetized by applying magnetic fields ranging from 1 to 9 T in the out-of-plane film direction along the beam path. The temperature was varied between 20 K and 300 K. The X-ray absorption data were acquired in the total electron yield (TEY) mode, which involves measuring the drain current from sample to ground. To ensure accuracy and eliminate any X-ray intensity fluctuations, a reference signal was recorded by monitoring the transmission of X-rays through a gold grid positioned along the beamline and upstream of the superconducting magnet. All spectra were subsequently normalized using this reference signal. This experimental setup allowed for precisely determining magnetic contributions at site-specific levels.

\section{Results and Discussion}

\subsection{Magnetometry Measurements}
 
To gain a deeper understanding of the magnetic behavior of TmIG thin films, we examine their magnetization response under different field orientations and temperatures. Figure \(\ref{Fig1}\)(c) presents the TmIG out-of-plane (OOP) and in-plane (IP) magnetization curves measured at two temperatures: 300 K and 60 K. At 300 K, an abrupt magnetic reversal is observed along the OOP field orientation. The sample exhibits a saturation magnetization of \(2\) \(\mu_B\)/f.u. and an extremely low coercivity of 0.002 T (see inset). In the IP configuration, the maximum observed is approximately \(1.0\) \(\mu_B\)/f.u., lower than that observed in the OOP configuration, indicating that stronger fields are required to saturate the sample along this direction. Furthermore, this reduction is accompanied by a decrease in the slope of the magnetization response at low magnetic fields. The combination of a smaller coercivity and sharper switching behavior suggests an easy axis of magnetization along the out-of-plane direction and the presence of a perpendicular magnetic anisotropy, consistent with the expected response for TmIG films under tensile strain \cite{Vilela20-Strain}.

At lower temperatures, an increase in magnetization noise is observed, primarily due to the reduced signal-to-noise ratio of the TmIG thin film relative to the paramagnetic GGG substrate. This behavior stems from the enhanced magnetic moment of Tm atoms at reduced temperatures, which, owing to their antiferromagnetic coupling with the Fe sublattice, results in an overall decrease in the net magnetization. In contrast to the hysteresis behavior observed at 300 K in the OOP configuration, a substantial increase in coercivity is observed at low temperatures, reaching approximately 0.0244 T, ten times more than at 300 K. Under these conditions, the sample exhibits a saturation magnetization of about \(1.5\) \(\mu_B\)/f.u.. In the IP configuration, the hysteresis loop displays a square shape at low fields, accompanied by a significantly higher coercivity of 0.11 T (see inset).

Figure \(\ref{Fig1}\)(d) illustrates the temperature dependence of the coercive field (\(H_c\)) in the OOP geometry. In rare-earth iron garnets such as TmIG, the evolution of \(H_c\) is governed by the interplay between magnetic anisotropy, sublattice interactions, and domain wall dynamics \cite{Uemura08-Double, Vértesy23-Coercive, Liang23-Thickness}. At high temperatures, \(H_c\) remains relatively low, consistent with reduced anisotropy. It increases as the temperature decreases, reaching a maximum near 50 K. It reflects enhanced anisotropy and stronger interaction between the Fe and Tm sublattices, owing to their respective increased magnetic moments and antiparallel alignment. This combination leads to reduced overall magnetization and, thus, an increase in the coercive fields \(H_c\). Similar trends were observed in other rare-earth garnet systems, particularly near their compensation temperatures, where anisotropy-driven transitions and fully compensated antiferromagnetism affect coercivity \cite{Vértesy23-Coercive, Liang23-Thickness}. It is worth noting the divergence of \(H_c\) near 50 K, highlighted by the dashed vertical line in Figure \(\ref{Fig1}\)(d). This behavior suggests a temperature-dependent modification in exchange interactions, likely influenced by the competition between the Tm and Fe sublattices, where their respective magnetic moments tend to cancel each other out, reducing the net moment. In orthoferrites, similar effects arise from the interplay between 3\(d\)–4\(f\) interactions and the temperature-dependent anisotropy of rare-earth ions, which modify the stability of the magnetic sublattices order \cite{Zhang16-Resolving}. In particular, thermal changes in the occupancy of the 4\(f\) states of Tm ions alter the energy balance between competing interactions, leading to magnetic phase transitions \cite{Staub17-Interplay}. A similar divergence in coercivity was reported in intermetallic ferrimagnets, such as DyCo\(_5\) and CoGd alloys, where \(H_c\) increases significantly near the magnetization compensation temperature \cite{Donges17-Magnetization, Fu21-Complex}. In these systems, the rise in coercivity results from the competition between sublattice magnetic moments, which approach equilibrium before reversing across the compensation point. Although a strict compensation temperature is not expected in bulk TmIG, the observed minimum in magnetization at 50 K coincides with the divergence in coercivity (see Figure \ref{Fig1}(e)). This compensation-like behavior at \(T_{\text{comp-like}}\) suggests that the interplay between anisotropy and exchange interactions plays a crucial role in governing the magnetic response of TmIG thin films, akin to the coercivity trends observed in intermetallic systems.

\begin{figure*}[ht]
    \centering
    \includegraphics[width=\textwidth]{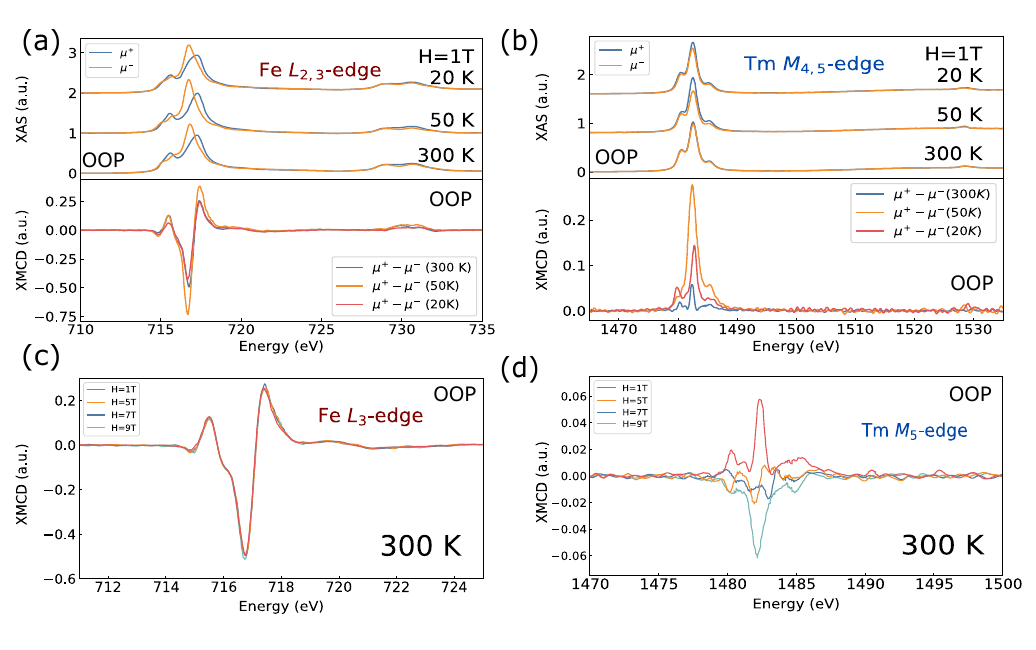}
    \centering
    \caption{(a) XAS and XMCD spectra at the Fe \(L_{2,3}\) edges for the TmIG thin film, measured under a 1 T magnetic field applied perpendicular to the film surface, at 20 K, 50K and 300 K. (b) XAS and XMCD spectra at the Tm \(M_{4,5}\) edges under the same conditions as in (a). (c) XMCD spectra at the Fe \(L_{3}\) and (d) Tm \(M_{5}\) edges at 300 K, highlighting the magnetic response under applied magnetic fields of up to 9 T.}
    \label{Fig2}
\end{figure*}

Figure \(\ref{Fig1}\)(e) presents the temperature dependence of the out-of-plane magnetization at 0.2 T, chosen to minimize the noise from the increasing paramagnetic contribution of the GGG substrate at higher fields. The magnetization exhibits a slight increase up to 200 K, followed by a steady decrease and the emergence of the aforementioned minimum near 50 K, coinciding with the coercivity divergence observed in Figure \(\ref{Fig1}(d)\). As previously mentioned, the reduction in overall magnetization due to the enhanced antiferromagnetic coupling between the Fe and Tm sublattices, as well as anisotropic effects, leads to an increase in coercivity. This divergence in \(H_c\) is a signature of its inverse dependence on magnetization near the compensation-like point, which scales as \(H_c \propto 1/M\). The observation of \(T_{\text{comp-like}}\) behavior is particularly noteworthy, as such a feature is not expected in bulk TmIG \cite{Geller65-Magnetic}. A direct comparison with earlier studies underscores the distinct nature of our findings. Wang et al. \cite{Wang19-Preparation} reported that TmIG thin films grown by pulsed laser deposition (PLD) on GGG substrates do not exhibit a compensation-like temperature, despite high crystalline quality and smooth interfaces. Sharma et al. \cite{Sharma23-Magnetic}, in turn, observed a compensation temperature near 15 K in sol-gel-based TmIG films deposited on oxidized silicon, which differ in substrate, thickness, and crystallinity. Notably, our 30-nm-thick TmIG film exhibits a compensation-like behavior around 50 K, significantly higher than that observed by Sharma et al. \cite{Sharma23-Magnetic}. This shift is likely linked to the tensile strain and improved crystallinity resulting from the 900 °C annealing, which alters the magnetic anisotropy and exchange interactions between the Fe and Tm sublattices. The emergence of perpendicular magnetic anisotropy (PMA) in our film appears to correlate with this behavior, reinforcing the notion that strain and processing conditions critically influence the compensation phenomena in TmIG systems.

\subsection{X-ray absorption spectroscopy}

\begin{figure*}[ht]
    \centering
    \includegraphics[width=\textwidth]{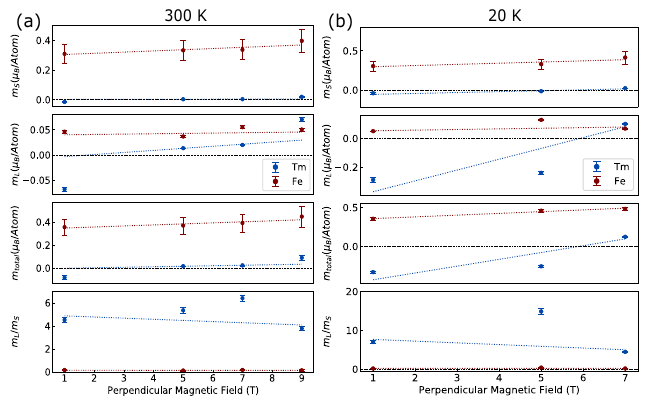}
    \centering
    \caption{Field dependence of the orbital (\(m_L\)), spin (\(m_S\)), total magnetic moments, and \(m_L/m_S\) ratio for Fe and Tm sublattices at 300 K (a) and 20 K (b). Dashed lines represent linear fits to the measured data.}
    \label{Fig3}
\end{figure*}

\begin{figure*}[ht]
    \centering
    \includegraphics[width=\textwidth]{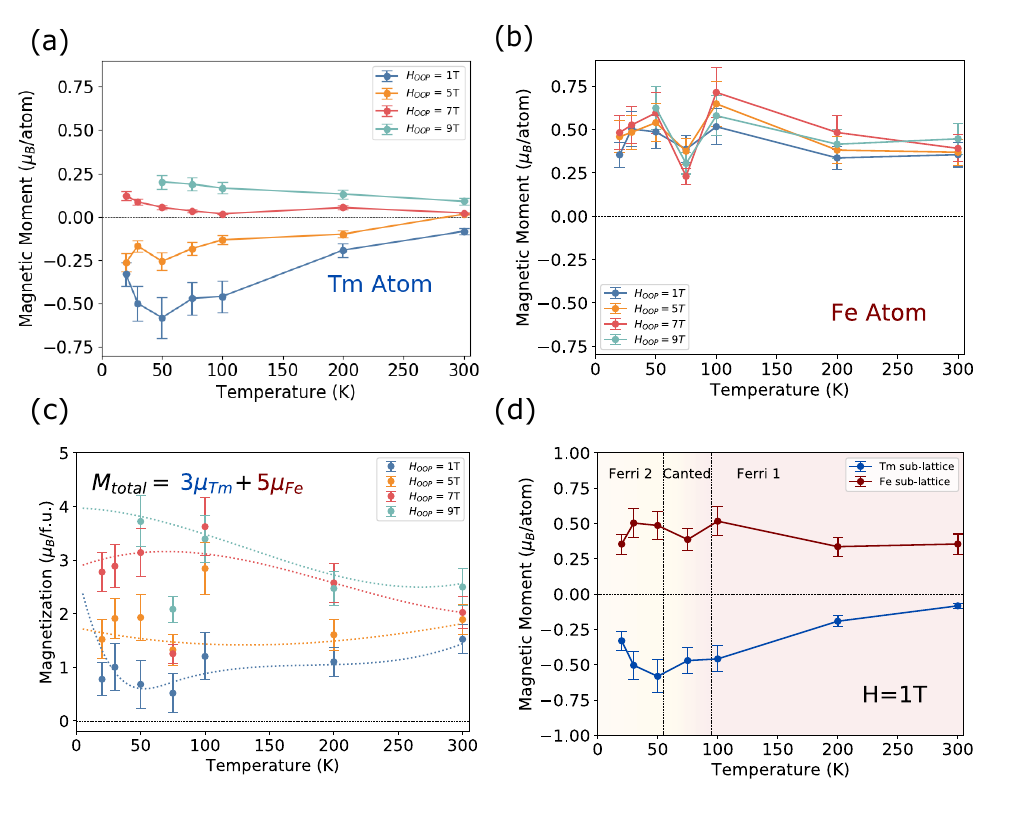}
    \centering
    \caption{(a) Magnetic moment per atom of the Tm sublattice as a function of temperature, indicating a strong field dependence and a transition in projection at higher fields. (b) Evolution of the total magnetic moment per atom of the net Fe sublattice, showing changes in the perpendicular projection below 100 K, with minimal influence from the applied magnetic field. (c) Total magnetization per formula unit (\(M_{\text{total}} = 3\mu_{\text{Tm}} + 5\mu_{\text{Fe}}\)) as a function of temperature for different magnetic fields, highlighting a magnetization minimum near 50 K for 1 T and its suppression at higher fields. (d) Detailed projections of the \(z\)-axis moments for Tm and Fe sublattices, revealing three distinct magnetic regimes (Ferri 1, Canted, and Ferri 2).}
    \label{Fig4}
\end{figure*} 

To elucidate the element-specific magnetic behavior of Fe and Tm in TmIG thin films, we performed X-ray absorption spectroscopy (XAS) and X-ray magnetic circular dichroism (XMCD) measurements under various magnetic fields and temperatures. Figure \(\ref{Fig2}\)(a) presents the XAS and XMCD spectra at the Fe \(L_{2,3}\) edges under an OPP 1 T magnetic field. The XMCD spectra reveal the distinct contributions of the tetrahedral and octahedral Fe sublattices, which are antiferromagnetically coupled, as expected for rare-earth iron garnets \cite{Quindeau16-Heterostructure, Vasili17-Direct, Tripathi2018}. The \(L_3\) edge region exhibits two positive peaks attributed to the octahedral (\(Oh\)) Fe\(^{3+}\) sublattice and a strong negative peak corresponding to the tetrahedral (\(Td\)) sublattice, reinforcing their antiferromagnetic alignment \cite{Tripathi2018}. In rare-earth iron garnets with magnetic compensation, such as DyIG, GdIG, and TbIG, the XMCD spectra often show a sign inversion at low temperatures due to the reversal of the tetrahedral and octahedral Fe sublattice moments. However, as observed in Figure \(\ref{Fig2}\)(a), no such inversion occurs in TmIG, neither at 50 K nor at 20 K. This suggests that the Tm sublattice plays a distinct role in modifying the Fe sublattice interactions, preventing a complete reversal of the Fe moments. The XMCD spectrum of Tm at the \(M_{4,5}\) edges (Figure \(\ref{Fig2}\)(b)) further confirms this behavior, where the signal intensity increases with decreasing temperature but does not invert, distinguishing TmIG from other RIGs with compensation temperature. In Figure \(\ref{Fig2}\)(b), an increase in the XMCD signal at 50 K compared to 300 K is observed, consistent with the expected enhancement of the total Tm moment at lower temperatures. Nevertheless, at 20 K, the XMCD intensity is lower than at 50 K, indicating a reduction in the projected Tm moment along the applied field direction.  

Figure \(\ref{Fig2}\)(c) shows the evolution of the XMCD spectra at the Fe \(L_3\) edge, while Figure \(\ref{Fig2}\)(d) presents the corresponding spectra at the Tm \(M_5\) edge under varying magnetic fields at 300 K. While the Fe spectra don't present important changes, a reduction in the XMCD intensity at the Tm \(M_5\) edge indicates a decrease in the projection of the Tm moment along the perpendicular direction as the magnetic field increases. For magnetic fields exceeding 5 T, the XMCD results suggest a reorientation of the Tm sublattice towards the magnetic field, altering its alignment with the Fe sublattices. We refer to this behavior as Tm spin-flip, which highlights the weaker coupling between the  Tm and Fe sublattices compared to the strong \(Fe_{d}-Fe_{a}\) superexchange interaction (a sketch representing the individual Fe and Tm sites is shown in Figure \ref{Fig1}(a)). \cite{Bayarra-Tuning}. A similar effect was reported in ErIG, where strong fields drive a transition to a canted phase \cite{Strohm12-Element}. The reorientation reflects the complex interplay of exchange interactions and anisotropy in TmIG, as well as the intricate spin-flip observed for higher fields. This work may open new possibilities to system engineering to reduce the critical magnetic field required for the spin-flip.  
    
Thereafter, we present the results for the orbital and spin magnetic moments extracted using the XMCD sum rules \cite{Teramura96-Effect, Tripathi2018}, with further details provided in the Supplementary Information. Figures \(\ref{Fig3}\)(a) and \(\ref{Fig3}\)(b) illustrate the evolution of the spin (\(m_S\)), orbital (\(m_L\)), as well as (\(m_L\)/\(m_S\)) and total magnetic moment (\(m_{\text{total}} = m_S + m_L\)) for the Fe and Tm sublattices as a function of magnetic field at 300 K and 20 K, respectively. The Fe sublattice exhibits weak dependence on the applied field at both temperatures. For example, at 300 K, the total Fe moment increases from 0.35 \(\mu_B\)/atom at 1 T to 0.45 \(\mu_B\)/atom at 9 T, with the spin moment remaining nearly constant (~0.31–0.40 \(\mu_B\)/atom). In contrast, the Tm sublattice shows a markedly stronger dependence on the external field. At 300 K, the Tm total moment evolves from –0.08 \(\mu_B\)/atom (opposite to the applied field) at 1 T to 0.09 \(\mu_B\)/atom at 9 T—an inversion that confirms a spin-flip transition between 5 and 7 T. A similar trend is observed at 20 K. The Fe sublattice maintains a nearly stable spin moment across the field range (0.30–0.41 \(\mu_B\)/atom). The Tm sublattice, however, presents a striking variation: the total moment rises from –0.33 \(\mu_B\)/atom at 1 T to 0.12 \(\mu_B\)/atom at 9 T, reversing its sign and more than tripling in magnitude. These results highlight the weaker exchange coupling of Tm compared to Fe and the field sensitivity of the rare-earth sublattice. Furthermore, the orbital-to-spin moment ratio \(m_L/m_S\) for Tm increases substantially with decreasing temperature, from $\approx$ 4.5 at 300 K to nearly 7 at 20 K at 1 T. This pronounced enhancement indicates that spin-orbit coupling plays a progressively larger role in determining the Tm magnetic moment as temperature decreases. At lower temperatures, thermal agitation is reduced, which reduces the randomization of spin and orbital orientations. In rare-earth ions like Tm\(^{3+}\), the \(4f\) electrons are strongly localized and experience significant spin-orbit coupling. Under reduced thermal energy, the orbital component of the magnetic moment becomes more stabilized due to the anisotropic distribution of the \(4f\) charge cloud in the crystal field environment. This stabilization enhances the orbital contribution (\(m_L\)) relative to the spin component (\(m_S\)). As a result, the \(m_L/m_S\) ratio increases significantly at low temperatures, as observed in our measurements. This behavior has also been reported in other rare-earth garnets, where the orbital moment overtakes the spin contribution as the dominant component of the magnetic moment in cryogenic regimes \cite{Carreta20-Interfacial, Omar24-Room}.

Overall, these findings demonstrated that the Tm sublattice underwent a field-induced spin-flip transition, where its magnetic moments — initially aligned antiparallel to the Fe sublattices due to inter-sublattice exchange coupling — reoriented along the direction of the applied field above a critical magnetic field. This reorientation, driven by the competition between the Zeeman energy and exchange interactions, resembled phenomena reported in other rare-earth oxides and offered promising mechanisms for magnetization control \cite{Cao14-Magnetization}. Combined with the enhanced orbital contribution at low temperatures, these results emphasized the tunability of TmIG thin films and their potential for spintronic technologies that require controllable and energy-efficient magnetic switching.

\subsection{Compensation-like transition and sublattice interactions}

To elucidate the temperature-dependent magnetic behavior and inter-sublattice coupling in TmIG, we systematically examined the evolution of the atomic magnetic moments of Tm and Fe across a range of magnetic fields. Figure \(\ref{Fig4}\)(a) presents the temperature dependence of the Tm total atomic moment for different magnetic fields. For lower fields (e.g., 1 T), the total Tm moment (\(m_{total}\)) decreases as the temperature drops from room temperature (300 K) to approximately 50 K, reaching a minimum of -0.6 $\mu_B$/atom  and then again increasing to -0.3 $\mu_B$/atom at 20 K \cite{Omar24-Room}. When increasing the field up to 5 T, the behavior remains similar, though the moment amplitude becomes smaller and exhibits only slight changes at low temperatures. At higher fields (\(\geq 7\) T), the projection of the Tm moment along the \(z\)-axis becomes positive, aligning with both the net Fe sublattice moment and the applied magnetic field over the entire temperature range reaching a maximum value near +0.2 $\mu_B$/atom at 20K. Notably, at 9 T, there is a significant increase in the \(z\)-axis projection of the Tm total atomic moment, almost doubling its value compared to 7 T, suggesting a trend toward collinear alignment of the Tm moments with the external field.

Figure \ref{Fig4}(b) displays the temperature evolution of the Fe atomic magnetic moment per atom, calculated for different magnetic fields. Upon cooling from 300 K to approximately 100 K, the Fe moment increases for all field values, reaching maximum values ranging from 0.516 $\mu_B$ (1 T) to 0.714 $\mu_B$ (7 T). Below this temperature range, the Fe moment decreases significantly, reaching values as low as 0.230 $\mu_B$ at 7 T. This corresponds to a reduction of up to $\approx$ 68\%, highlighting the strong suppression of the Fe sublattice magnetization at low temperatures. The decrease becomes more pronounced as the field increases. This temperature dependence roughly mirrors the behavior of the Tm sublattice, as shown in Figure~\ref{Fig4}(a), and together, they lead to the minimum total magnetization observed in the SQUID measurements. This compensation-like temperature, $T_{\text{comp-like}}$, associated with the partial cancellation of sublattice moments, occurs around 50 K at 1 T, as shown in Figure~\ref{Fig1}(d).

In Figure \ref{Fig4}(c), the total magnetic moment per formula unit, considering the Tm\(_3\)Fe\(_5\)O\(_{12}\) composition and calculated as \(M_{\text{total}} = 3\mu_{\text{Tm}} + 5\mu_{\text{Fe}}\), is plotted as a function of temperature for different applied magnetic fields. At 1 T, a compensation-like behavior emerges around 50 K, resulting from the increasing antiparallel contribution of the Tm sublattice to that of Fe, which suppresses the net magnetic moment. This occurs because, at 1 T and 5 T, the magnetic moment of the Tm atom projects along the \(z\)-axis in a direction opposite to the applied field and opposite to the Fe sublattice, as evidenced by its negative value. As the field increases, this suppression gradually weakens, and the \(z\)-axis projection of the Tm moment becomes positive. For fields above 7 T, the minimum associated with the compensation-like behavior disappears entirely, indicating that the Tm moment aligns with the Fe sublattice. This suggests a transition towards a more collinear spin configuration under strong magnetic fields.

Figure \ref{Fig4}(d) provides further insight into the \(z\)-axis projections of the Fe and Tm sublattices, identifying three distinct magnetic regimes at 1 T. Above 100 K, the system is in a collinear ferrimagnetic configuration (Ferri 1), where the Fe and Tm sublattice moments are antiparallel and aligned along the field direction. Between 100 K and 50 K, the system remains ferrimagnetic, but the Tm projection increases in negative values while the Fe projection decreases, indicating a reconfiguration of the spin structure. Although our XMCD measurements do not directly reveal a non-collinear structure, we propose the emergence of a slightly canted configuration in this temperature range, supported by the observed changes in the sublattice projections. Similar behavior has been reported in other rare-earth iron garnets, such as ErIG and TbIG, where temperature-induced canting near \( T_{\text{comp}} \) has been associated with a competition between exchange coupling and local anisotropy effects \cite{Lahoubi14-Anomalous, Strohm12-Element, Tomasello22-Origin}. XMCD, being element-specific and sensitive to the projection of the magnetic moment along the \(z\)-axis, reveals that the Tm moment reverses its sign around 50 K at low fields, an indication of spin reorientation that aligns with previous observations of umbrella-like or canted spin states in this family of materials. Below 50 K, the system enters a second collinear regime (Ferri 2), where the \(z\)-axis projections of both Fe and Tm decrease further, and the Tm sublattice continues to oppose the Fe sublattice. These findings underscore the complex interplay between exchange interactions, spin-orbit coupling, and rare-earth anisotropy in shaping the temperature- and field-dependent spin configurations of TmIG thin films.

\section{Conclusion}

In conclusion, this study provides a comprehensive analysis of the magnetic behavior of TmIG thin films, offering new insights into the interplay between exchange interactions, compensation temperature, and spin-flip switching. The observed magnetization minimum near 50 K under moderate magnetic fields, absent in bulk TmIG, indicates a temperature-dependent competition of the Tm and Fe sublattices, shaped by spin-orbit coupling and local anisotropy variations. This compensation-like temperature  \(T_{\text{comp-like}}\) is reinforced by the dependence of the coercive field as a function of the temperature, which diverges around the \(T_{\text{comp-like}}\) point. The emergence of \(T_{\text{comp-like}}\) in TmIG thin films is particularly noteworthy and may provide technologically relevant functionalities, such as low-power spin manipulation and ultrafast magnetization reversal \cite{Deb18-Controlling, Zhao16-Origin}.

In addition, the field-induced spin flip of the Tm sublattice observed in this study introduces another important fundamental physical understanding of the magnetic reversal of ferrimagnetically coupled sublattices. As the field increases, the Tm magnetic moments switch toward the external field, while the Fe magnetic moments remain unchanged, reflecting the weaker coupling of Tm moments compared to the strong Fe-Fe exchange interaction. This phenomenon can be exploited by designing materials with tunable magnetic properties to reduce the critical magnetic fields where the Tm spin-flip occurs. The ability to induce spin reorientation in TmIG via moderate external fields could be used to further investigation on this relevant technological material, with implications for both magnonic and spintronic applications. 

\begin{acknowledgement}

The authors thank the National Institute of Science and Technology for Spintronics and Advanced Magnetic Nanostructures (INCT-SpinNanoMag) and the National Council for Scientific and Technological Development (CNPq, grant 168483/2023-8) for financial support. This work was also supported by the São Paulo Research Foundation (FAPESP, grant 2017/10581-1) and CNPq (grants 312762/2021-6 and 421070/2023-4). Additional support was provided by the Air Force Office of Sponsored Research (FA9550-23-1-0004 DEF), the National Science Foundation (NSF-DMR 2218550 and 1231319), and the Army Research Office (W911NF-20-2-0061, DURIP W911NF-20-1-0074).

\end{acknowledgement}




\bibliography{reference}
\end{document}